\documentclass{sig-alternate-05-2015}
\usepackage{color}
\usepackage{algorithm}
\usepackage{algpseudocode}
\usepackage{cite}

\usepackage{array}
\usepackage{eqparbox}
\usepackage{stfloats}
\usepackage{url}
\usepackage{color}
\usepackage{multirow}
\usepackage{amsmath,amsfonts}
\usepackage{amssymb}
\usepackage[colorlinks]{hyperref}
\usepackage{subfigure}
\usepackage{caption2}
\usepackage{graphicx}
\usepackage{tikz}

\newcommand{\pgftextcircled}[1]{
    \setbox0=\hbox{#1}%
    \dimen0\wd0%
    \divide\dimen0 by 2%
    \begin{tikzpicture}[baseline=(a.base)]%
        \useasboundingbox (-\the\dimen0,0pt) rectangle (\the\dimen0,1pt);
        \node[circle,draw,outer sep=0pt,inner sep=0.1ex] (a) {#1};
    \end{tikzpicture}
}

\newcommand{\pgftextcircledblk}[1]{
    \setbox0=\hbox{#1}%
    \dimen0\wd0%
    \divide\dimen0 by 2%
    \begin{tikzpicture}[baseline=(a.base)]%
        \useasboundingbox (-\the\dimen0,0pt) rectangle (\the\dimen0,1pt);
        \node[circle,draw,outer sep=0pt,inner sep=0.1ex,fill=blue] (a) {#1};
    \end{tikzpicture}
}

\def\t0{{t_0}}

\def\N{{\mathbb N}}
\def\R{{\mathbb R}}        

\def\t0{{t_0}}

\def\prox{{\rm prox}}

  \newtheorem{remark}{Remark}
\begin{document}




\CopyrightYear{2016}
\setcopyright{acmcopyright}
\conferenceinfo{MM '16,}{October 15-19, 2016, Amsterdam, Netherlands}
\isbn{978-1-4503-3603-1/16/10}\acmPrice{\$15.00}
\doi{http://dx.doi.org/10.1145/2964284.2964298}

\clubpenalty=10000
\widowpenalty = 10000

\title{Parsimonious Mixed-Effects HodgeRank for Crowdsourced Preference Aggregation}

\numberofauthors{1}
\author{
    \alignauthor
    Qianqian Xu$^{1,2}$, Jiechao Xiong$^{3}$, Xiaochun Cao$^{1\thanks{Corresponding author.}}$, Yuan Yao$^{3^{\ast}}$\\
    \affaddr{$^1$ State Key Laboratory of Information Security, Institute of Information Engineering, Chinese Academy of Sciences, Beijing 100093, china}\\
    \affaddr{$^2$ BICMR, Peking University, Beijing 100871, China}\\
    \affaddr{$^3$ School of Mathematical Sciences, BICMR-LMAM-LMEQF-LMP, Peking University, Beijing 100871, China}\\
    \email{\{xuqianqian, caoxiaochun\}@iie.ac.cn\quad xiongjiechao@pku.edu.cn\quad yuany@math.pku.edu.cn}
}

\maketitle
\begin{abstract}

In crowdsourced preference aggregation, it is often assumed that all the annotators are subject to a common preference or utility function which generates their comparison behaviors in experiments.
However, in reality annotators are subject to variations due to multi-criteria, abnormal, or a mixture of such behaviors. In this paper, we propose a parsimonious mixed-effects model based on HodgeRank, which takes into account both the fixed effect that the majority of annotators follows a common linear utility model, and the random effect that a small subset of annotators might deviate from the common significantly and exhibits strongly personalized preferences. HodgeRank has been successfully applied to subjective quality evaluation of multimedia and resolves pairwise crowdsourced ranking data into a global consensus ranking and cyclic conflicts of interests. As an extension, our proposed methodology further explores the conflicts of interests through the random effect in annotator specific variations. The key algorithm in this paper establishes a dynamic path from the common utility to individual variations, with different levels of parsimony or sparsity on personalization, based on newly developed Linearized Bregman Algorithms with Inverse Scale Space method. Finally the validity of the methodology are supported
by experiments with both simulated examples and three real-world crowdsourcing datasets, which shows that our proposed method exhibits better performance (i.e. smaller test error) compared with HodgeRank due to its parsimonious property.
\end{abstract}

%
%
%

%
%

%
%
\printccsdesc



\section{Introduction}
\begin{figure}[t]
 \begin{center}
\includegraphics[width=\linewidth]{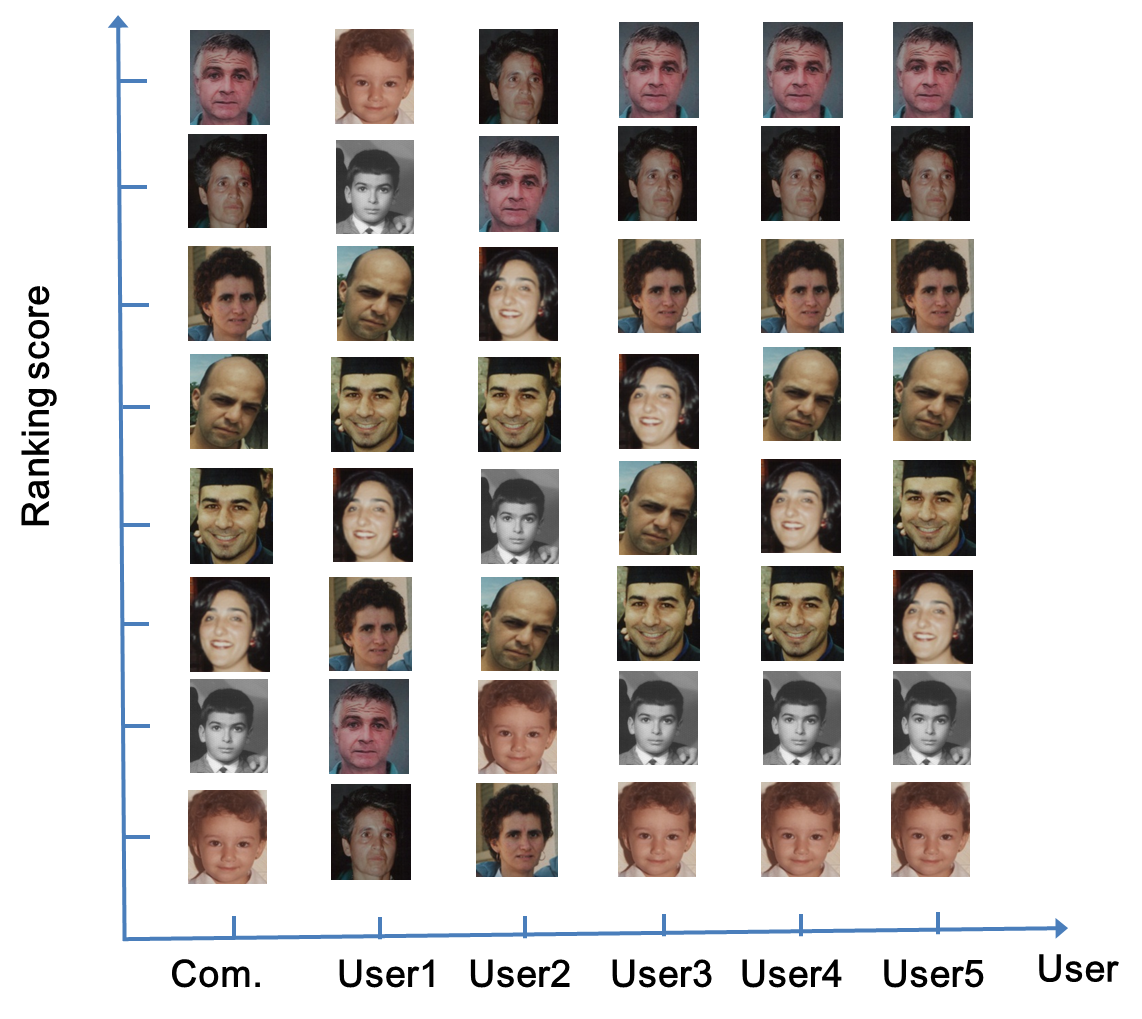}
  \caption{An illustrative example on estimation of human ages, with the fixed effect of common ranking and random effects of user's personalized ranking. The faces in the first column are ordered according to a common ranking score aggregated from crowdsourced pairwise comparisons, while other columns are according to personalized rankings of different users. The ground truth ages in the first column, in a top-down order, are 46, 51, 36, 30, 25, 23, 10, and 2, respectively.} \label{introimage}
\end{center} \label{fig:age0}
\end{figure}

With the Internet and its associated explosive growth of information, individuals today in the world are facing with the
rapid expansion of multiple choices (e.g., which book to buy, which hotel to book, etc.). Inferring user's preference or utility over a set of alternatives has thus become an important
issue. Among various methods to infer user viewpoint/preference, crowdsourcing technology is becoming a new paradigm, which collects voting data from a large crowd or population on Internet and pursue some statistical preference aggregations. For example, the following platforms are frequently used by researchers to crowdsource voting data of participants: \href{https://www.mturk.com}{MTurk}, \href{http://www.innocentive.com/}{InnoCentive}, \href{http://crowdflower.com/}{CrowdFlower},~\href{http://www.crowdrank.net/}{CrowdRank}, and \href{http://www.allourideas.org/}{AllOurIdeas}, etc. A typical and perhaps the simplest scenario is the pairwise comparison experiment. Specifically,
there are a set of items to rank, and participants are asked to choose between various pairs among these items; the goal is to aggregate these pairwise comparisons
into a global consensus ranking that summarizes the preference of all users. We have seen that researchers exploit such a paradigm to evaluate the quality of multimedia content \cite{MM09,MM13}, predict image/video interestingness \cite{fu2014interestingness}, estimate ages from face pictures \cite{fu2015robust}, and rank taste of food \cite{jamieson2011active} etc.

However, different individuals might very well have distinct preferences, such that participants of the crowdsourced experiments might vote under different criteria or conditions. It might be misleading to merely look at a global consensus while ignoring personal diversity. For example Fig.\ref{fig:age0} shows an age estimation from photos that will be discussed in detail later in this paper. The majority is not always correct, as the common ranking by mistake thinks 46 older than 51! Moreover, even though {\tt User2} is largely deviated from the common ranking, it is noticeable that he/she makes correct judgements between the faces of year 46 and 51. So if one is looking for features that correctly predict the ages of these two particular faces, {\tt User2} is a better consultant than the majority. Particularly, {\tt User1} is clearly an adversarial voter, whose personalized ranking is largely against the common ranking reflecting the majority, so should be removed from a preference aggregation procedure.

Moreover, in crowdsourcing experiments, the participants are distributed over the Internet with a diverse environment. Even they might share the same preference or utility function in making choices, they might suffer various disturbances during the experiments. For example, i) one typically clicks one side more often than another. As some pairs are highly confusing or annotators get too tired, in these cases, some annotators tend to click one side hoping to simply raise their record to receive more payment; while for pairs with substantial differences, they click as usual. ii) some extremely careless annotators, or robots pretending to be human annotators, actually do not look at the instances and click one side all the time to quickly receive payment for work. Such a kind of behavior is called the annotator's position bias which has been studied in \cite{day1969position,xu2016}.

These examples above suggest us that we have to take into account of user or annotator specific variations in a crowdsourced preference aggregation task.
As the classical social choice theory \cite{Arrow51} points out, preference aggregation toward a global consensus is doomed to meet the conflicts of interests. \emph{What is a suitable way to quantitatively analyze the conflicts of interests}?

In this paper, we pursue the Hodge-theoretic approach by \cite{Hodge} which decomposes the pairwise comparison data into three orthogonal components: the global consensus ranking, the local inconsistency as triangular cycles, and the global inconsistency as harmonic cycles. The latter two are both cycles, collectively decoding all the conflicts of interests in the data. Instead of the merely extracting from the data the global ranking component, often called \emph{HodgeRank} which has been introduced into the quality assessment of multimedia by \cite{tmm12}, we extend it here by including some annotator-specific random effects to further decompose the cycles. To decipher the sources of the conflicts of interests, we mainly consider two types of annotator-specific variations: annotator's personalized preference deviations from the common ranking which characterize multi-criteria in data, and annotator's position bias which deteriorates the quality of data. This results in a linear mixed-effects extension of HodgeRank, called Mixed-Effects HodgeRank here.


To initiate a task of crowdsourced preference aggregation, we usually assume the majority of participants share a common preference interest and behave rationally, while deviations from that exist but are sparse. So a parsimonious model is assumed in this paper, with sparsity structure on personalized preference deviations and position biases. Due to the unknown amount of such sparse random effects in reality, it is natural to pursue a family of parsimonious models at a variety levels of sparsity. Algorithmically we adopt the Linearized Bregman Iteration, which is a simple iterative procedure generating a sequence of parsimonious models, evolving from the common global ranking in HodgeRank, to annotator's personalized ranking till a full model. Fig.\ref{fig:age0} is in fact a result of our algorithm. As the algorithm iterates, typically the abnormal annotators with large preference deviations and/or position biases appear early, and the annotators who behave normally appear at a later stage. In practice when the number of participants is large and sample size is relatively small, early stopping regularization is needed to prevent the overfitting in full model.

Equipped with such a new scheme, given a set of entities, we choose a
set of entity pairs and ask Internet crowds which entity is more preferable
in each pair. Based on the feedback we not only can derive the common preference on population-level, but also can estimate rapidly an annotator's large preference/utility deviation in an individual-level, and an abnormal annotator's position bias. Individual preference deviations from the population common ranking are helpful to understand different criteria among annotators when they judges, and especially to monitor the adversarial users. On the other hand, annotator's position bias is a helpful tool to monitor the quality of his/her voting data, through the mixing behavior that the annotator simply clicks one side of the pair in comparisons without paying attention to their contents. Such a statistical mixed-effects framework simultaneously considers both the fixed effect of common ranking as the HodgeRank and the random effects of annotator-specific variations, which, up to the author's knowledge, has not been seen in literature.

As a summary, our main contributions in this new framework are highlighted as follows:

\begin{itemize}
\item[(A)] A linear mixed-effects extension of HodgeRank including both the fixed effect of common ranking, and the random effects of annotator's preference deviation with position bias;
\item[(B)] A path of parsimonious estimates of the preference deviation and position bias at different sparsity levels, based on Linearized Bregman Iterations.
\end{itemize}

The remainder of this paper is organized as follows. Sec.\ref{sec:relatedwork} contains a review of related works. Then we systematically introduce the methodology for parsimonious mixed-effects HodgeRank estimation in Sec.\ref{sec:methodology}. Extensive experimental validation based on one simulated and three real-world crowdsourced datasets are demonstrated in Sec.\ref{sec:experiment}. Finally, Sec.\ref{sec:conclusion} presents the conclusive remarks.

\section{Related Work} \label{sec:relatedwork}

\subsection{Statistic Ranking Aggregation}

Statistical preference aggregation, in particular ranking or rating from pairwise comparisons,
is a classical problem which can be traced back to the $18^{th}$ century.
Various algorithms have been studied for this problem, including maximum
likelihood under a Bradley-Terry model
assumption, rank centrality (PageRank/MC3) \cite{negahban2012,dwork2001rank}, HodgeRank \cite{Hodge}
, and a pairwise variant of Borda
count \cite{de1781memoire} among others. In \cite{ICML14}, it shows that under a natural statistical model, where pairwise comparisons
are drawn randomly and independently from some
underlying probability distribution, the rank centrality (PageRank) and
HodgeRank algorithms both converge to an optimal ranking under
a ``time-reversibility" condition. However, PageRank is only able to aggregate the pairwise comparisons
into a global ranking over the items. HodgeRank not only provides us a mean to determine a global ranking under
various statistical models, but also measures the inconsistency of the global ranking obtained.

HodgeRank, as an application of combinatorial Hodge theory to the preference or rank aggregation problem from pairwise comparison data, was first introduced in~\cite{Hodge}, inspiring a series of studies in statistical ranking~\cite{hodge_l1,osting2013enhanced,osting2016analysis}, game theory~\cite{Parrilo11_gameflow}, and computer vision~\cite{Yuan09_hodge}, {etc}. It is a general framework to decompose paired comparison data on graphs, possibly imbalanced (where different video pairs may receive different number of comparisons) and incomplete (where every participant may only give partial comparisons), into three orthogonal components. In these components HodgeRank not only provides us a mean to determine a global ranking from paired comparison data under various statistical models (e.g., Uniform, Thurstone-Mosteller, Bradley-Terry, and Angular Transform), but also measures the inconsistency of the global ranking obtained. The inconsistency shows the validity of the ranking obtained and can be further studied in terms of its geometric scale, namely whether the inconsistency in the ranking data arises locally or globally. Local inconsistency can be fully characterized by triangular cycles, while global inconsistency involves cycles consisting nodes more than three, which may arise due to data incompleteness and once presented with a large component indicates some serious conflicts in ranking data. However through random graphs, we can efficiently control global inconsistency.

However, all of these methods have a major drawback: they aim to find one
ranking thus cannot
model the discrepancies across users. Therefore, in recent years, personalized ranking methods arise to fill in this gap.
This task can be seen as rank aggregation
analog to the standard collaborative filtering (CF) problem.
There
have been many CF algorithms, including Bayesian networks,
clustering models, and latent semantic models, etc. Recent algorithms
 for collaborative filtering are mostly based on matrix factorization \cite{salakhutdinov2008bayesian,rennie2005fast}.
  The key idea behind them is to find a low rank user rating matrix that best approximates the observed ratings. Most recently, the application of the nuclear norm approach to CF was first proposed by \cite{yi2013inferring}, which shows good empirical evidence for using such a nuclear
norm regularized based approach. The key difference between our
study and the low rank matrix collaborative filtering algorithms is that we assume the majority of voters share a fixed effect of common ranking while some annotators might deviate from that significantly. Such parsimonious model from population to individual is a natural fit for crowdsourcing scenarios.

\subsection{Linearized Bregman Iteration (LBI)}
Linearized Bregman Iteration (LBI) was firstly introduced in \cite{YODG08} in the literature of variational imaging and compressive sensing. It is well-known that LASSO estimators are always biased \cite{Fan2001}. On the other hand, \cite{OBG+05} notices that Bregman iteration may reduce bias, also known as contrast loss, in the context of Total Variation
image denoising. Now LBI can be viewed as a discretization of differential equations (inclusions), called \emph{Inverse Scale Spaces}, 
which may produce unbiased estimators under nearly the same model selection consistency conditions as LASSO \cite{osher2014}. 

Beyond such a theoretical attraction, LBI is an extremely simple algorithm which combines an iterative gradient descent algorithm together with a soft thresholding. It is different to the well-known iterative soft-thresholding algorithm (ISTA) (e.g., \cite{Donoho95,fista} and references therein) which converges to the biased LASSO solution. To tune the regularization parameter in noisy settings, one needs to run ISTA with many different thresholding parameters and chooses the best among them; in contrast, LBI only runs in a single path and regularization is achieved by early stopping like boosting algorithms \cite{osher2014}, which may save the computational cost greatly and thus suitable for large scale implementation, e.g., distributive computation \cite{LBI_decentral}. 

\section{Methodology}
\label{sec:methodology}

In this section, we systematically introduce the methodology
for parsimonious mixed-effects HodgeRank estimation. Specifically,
we first start from extending the HodgeRank to a linear mixed-effect model.
Then we present a simple iterative algorithm called Linearized Bregman Iterations to generate paths of parsimonious models at different sparsity levels.
Early stopping regularization is discussed in the end.

\subsection{Mixed-Effects HodgeRank on Graphs}

In crowdsourced pairwise comparison experiments, Let $V = \{1,2,\dots,n\}$ be the set of nodes and $E = \{(u,i,j): i,j\in V, u \in U\}$ be the set of edges, where $U$ is the set of all annotators. Suppose the pairwise ranking data is given as $y: E\rightarrow R$. $y_{ij}^u>0$ means $u$ prefers $i$ to $j$ and $y_{ij}^{u}\leq 0$ otherwise. The magnitude of $y_{ij}^u$ can represent the degree of preference and it varies in applications. The simplest setting is the binary choice, where
\begin{align}
y_{ij}^u=\left\{\begin{array}{cc}
                                                  1 & \mathrm{if} \ u \ \mathrm{prefers} \ i \ \mathrm{to} \ j, \\
                                                  -1 & \mathrm{otherwise}.
                                                \end{array}
                                                \right.
\label{eq:Y}
\end{align}

%
%

The general purpose of preference aggregation is to look for a global score $\theta\colon V\to \R$ such that
\begin{equation} \label{eq:ho_rank0}
\min_{\theta\in {\mathbb{R}}^{|V|}} \sum_{i,j,u} \omega_{ij}^u l(\theta_i - \theta_j, y_{ij}^u),
\end{equation}
where $l(x,y)\colon \R\times \R\to \R$ is a loss function, $\omega_{ij}^u$ denotes the confidence weights on $\{i,j\}$ made by rater $u$ (for simplicity, assumed to be $\omega_{ij}^u=1$ for the existing voting data), and $\theta_i$ ($\theta_j$) represents the global ranking score of item \emph{i} (\emph{j}, respectively). In HodgeRank, one benefits from the use of square loss $l(x,y)=(x-y)^2$ which leads to fast algorithms to find optimal global ranking $\theta$, which becomes one component of a general orthogonal decomposition of paired comparison data \cite{Hodge}, i.e.
\[ y= global\ ranking \oplus cycles, \]
where the component \emph{cycles} can be further decomposed into
\[cycles = local\ cycles \oplus global\ cycles. \]

Local cycles are triangular cycles, e.g. $i \succ j \succ k\succ i $; while global cycles, also called harmonic cycles, are loops involving nodes more than three (e.g. $i \succ j \succ k \succ...\succ i$) and typically traversing all nodes in the graph. These cycles may arise due to conflicts of interests in ranking data. Therefore to analyze the statistical models of cycles is crucial to understand the conflicts of interests.

In crowdsourcing scenarios, the conflicts of interests are mainly due to two kinds of sources: the multi-criteria adopted by different annotators when they compare items in $V$; the abnormal behavior of annotators in the experiments, e.g. simply clicking one side of the pair when they got bored, tired, or distracted. To quantitatively characterize these effects, we propose the following model of cycles
\[ cycles = personalized\ ranking + position\ bias + noise. \]

To be specific, together with the global ranking component in HodgeRank, we consider the following linear mixed effects model for annotator's pairwise ranking:
\begin{equation} \label{eq:linear}
y_{ij}^u = (\theta_i+\delta_i^u) - (\theta_j+\delta_j^u) + \gamma^u + \varepsilon_{ij}^u,
\end{equation}
where
\begin{itemize}
\item $\theta_i$ is the common global ranking score, as a fixed effect;
\item $\delta_i^u$ is the annotator's preference deviation from the common ranking $\theta_i$ such that $\theta_i^u:=\theta_i + \delta_i^u$ becomes annotator $u$'s personalized ranking score, as a random effect;
\item $\gamma^u$ is an annotator's position bias, which captures the careless behavior by clicking one side during the comparisons;
\item $\varepsilon_{ij}^u$ is the random noise which is assumed to be independent and identically distributed with zero mean and being bounded.
\end{itemize}

Putting in matrix form, \eqref{eq:linear} becomes
\begin{equation} \label{eq:linear-matrix}
y = d\theta + X\beta + \varepsilon,
\end{equation}
where $d\in \R^{|E| \times |V|}$ satisfies $d\theta(u,i,j) = \theta_i-\theta_j$, $\beta = [\delta, \gamma] \in R^{(|V|+1)|U|}$ and $X =[D,A]$, where $D \in \R^{|E| \times |V| |U|} $ satisfies $D\delta(u,i,j) = \delta^u_i-\delta^u_j$ and $A \in \R^{|E| \times |U|}$ satisfies $\gamma(u,i,j) = \gamma^u$.

Here $\theta$ is population-level parameter which indicates some common score on $V$.  In crowdsourcing studies, as the data vary greatly
across individual annotators, we thus allow each annotator to have personalized
parameters $\theta^u$. These personalized parameters can be obtained by adding some
random effects $\delta^u$ to the population parameter $\theta$, representing individual deviations from the population behavior. Besides, $\gamma^u$ measures an annotator's position bias, i.e. the tendency of $u$ always clicking one side in paired comparison experiments. Under the random design of pairwise comparison experiments, a candidate should be placed on the left or the right randomly, so the position should not affect the choice of a careful annotator. However, some annotator might get confused, tired or distracted in experiments, such that he/she always clicks one side during some periods in experiments. Such a type of position bias captures a kind of noise in data not included in the zero mean $\varepsilon$ and may severely deteriorates the quality of data. The remainder $\varepsilon_{ij}^u$ measures the random noise in sampling which is of zero mean and bounded (hence subgaussian).

\subsection{Parsimonious Paths with Linearized Bregman Iteration}

In crowdsourced preference aggregation scenarios with good controls, it is natural to assume a parsimonious model. In such a model, the majority of annotators carefully follows the common behavior governed by the fixed effect parameter $\theta$, while only a small set of annotators might have nonzero personalized deviations and abnormal behavior in position bias. This amounts to assume that parameter $\delta^u$ to be group sparse, i.e. $\delta_i^u$ vanishes for all $i$ simultaneously, and $\gamma^u$ to be sparse as well, i.e. zero for most of careful annotators. Such a sparsity pattern motivates us to consider the following penalty function with a mixture of LASSO ($L_1$) penalty on $\gamma$ and group LASSO penalty on $\delta$:

\begin{equation}\label{eq-penalty}
P(\beta) = \|\gamma\|_1 + \sum_{u}\|\delta^u\|_2, \beta=(\delta,\gamma).
\end{equation}

{\begin{remark}
Usually a normalization factor $\sqrt{n}$ is used before a group lasso penalty $\|\delta^u\|_2$, where $n$ is the group size of $\delta^u$. But here all the $\delta^u$ have the same group size, and $\|D^u\|_F = \sqrt{2}\|A^u\|_F$, so the column norm of $D^u$ is on average $\frac{\sqrt{2}}{\sqrt{n}}$ times of $\|A^u\|_F$, this basically cancels out the factor $\sqrt{n}$. So here we just use this simple formula.
\end{remark}

Following the square loss in HodgeRank, the Euclidean distance (mean square error) in $R^{E}$ can be used for the total loss function:
\begin{equation}\label{eq-loss}
L(\theta,\beta) = \frac{1}{2m}\|y - d\theta - X\beta\|_2^2.
\end{equation}

The following Linearized Bregman Iterations (LBI) give rise to a sequence of parsimonious (sparse) models:
\begin{subequations}\label{eq:lbi0}
\begin{align}
\theta^{k+1} & = \theta^k - \alpha \kappa  \nabla_\theta L(\theta^k,\beta^k) \label{eq:lbi0-a}\\
z^{k+1} & = z^k - \alpha \nabla_\beta L(\theta^k,\beta^k), \label{eq:lbi0-b}\\
 \beta^{k+1} &=\kappa \cdot {\prox}_{P}(z^{k+1}), \label{eq:lbi0-c}
\end{align}
\end{subequations}
where $\beta^0 = 0$, $\theta^0 = \arg \min_{\theta} L(\theta,0)$, and variable $z$ is an auxiliary parameter used for gradient descent, $z = \rho+\beta/\kappa, \rho \in \partial P(\beta)$. Besides,
the proximal map associated with the penalty function $P$ is given by
\[ \prox_P(z) = \arg\min_{v \in R^{(|V|+1)|U|}} \left( \frac{1}{2}\| v - z\|^2 + P(z) \right ). \]

The Linearized Bregman Iteration \eqref{eq:lbi0} generates a path of global ranking score estimators $\theta^k$ and sparse estimators for preference deviation and position bias, $\beta^k=(\delta^k,\gamma^k)$. It starts from the common HodgeRank as $\theta^0 = \arg \min_{\theta} L(\theta,0)$, and evolves into parsimonious mixed effect models with different levels of sparsity until the full model, often overfitted. To avoid the overfitting, early stopping regularization is required to find an optimal tradeoff between the model complexity and in-sample error. For more details, we refer the readers to see \cite{osher2014} and references therein. In this paper, we find that cross validation works to find the early stopping time that will be discussed in Sec.\ref{sec:cv}.

The Linearized Bregman algorithm was firstly introduced in \cite{YODG08} extended from Bregman iteration \cite{OBG+05} as a scalable algorithm for large scale image restoration and compressed sensing. It has several advantages than the widely used LASSO-type convex regularizations. First of all, it is simpler than LASSO in generating the sparse regularization paths: instead of a parallel run of several optimization problem over a grid of regularization parameters, a single run of LBI generates the whole regularization path. LBI is thus desired in dealing with big problems. Moreover, it can be less biased than LASSO as if nonconvex regularizations \cite{Fan2001}. In fact, it is shown recently in \cite{osher2014} that as $\kappa\to \infty$ and $\alpha\to 0$, the limit dynamics of Linearized Bregman Iterations in sparse linear regression with LASSO ($L_1$) penalty may achieve the model selection consistency under nearly the same condition as LASSO yet return the unbiased Oracle estimator, while the LASSO estimator is well-known biased.



Here we give some remarks on the implementation details of the Linearized Bregman Iterations \eqref{eq:lbi0}.
\begin{itemize}
\item The parameter $\kappa$ determines the bias of the sparse estimators, a bigger $\kappa$ leading to the less biased ones. The parameter $\alpha$ is the step size which determines the precise of the path, with a large $\alpha$ rapidly traversing a coarse grained path. However one has to keep $\alpha \kappa$ small to avoid possible oscillations of the paths, e.g. $\alpha \kappa \|X^T X\|_2/m<2$. The default choice in this paper is $\alpha  = \frac{m}{\kappa\|X^TX\|_2}$ as a tradeoff between performance and computation cost.
\item The step \eqref{eq:lbi0-a} can also be replaced by $$\theta^{k+1}  = \arg\min_{\theta}L(\theta,\beta^k)$$ if it is easy to solve.
\item Now we turn to simplify the third step \eqref{eq:lbi0-c} with an explicit formula for the proximal map with the particular penalty function defined in Eq. (\ref{eq-penalty}). Recovering $\beta^{k+1}$ from $z^{k+1}$ is equivalent to the following group shrinkage on each group component of $\beta$, i.e. $\gamma^u$ and $\delta^u$:
\begin{eqnarray}
\beta^{k+1} &=& \kappa \mathbf{Shrinkage}(z^{k+1})\\
&\triangleq &\left\{
\begin{aligned}
\delta^{u,k+1} &= \kappa\max(0,1-1/\|z_{\delta^u}\|_2)z_{\delta^u} \nonumber \\
\gamma^{u,k+1} &= \kappa\max(0,1-1/|z_{\gamma^u}|)z_{\gamma^u} \nonumber \\
\end{aligned}
\right.
\end{eqnarray}
\end{itemize}


Now we are ready to give the following Linearized Bregman Algorithm for our Mixed-Effects HodgeRank as
\begin{algorithm}
\caption{LBI for ME-HodgeRank}\label{alg-LBI-LME}
\textbf{Input:} Data $(d,X,y)$, damping factor $\kappa$, step size $\alpha$.\\
\textbf{Initialize:} $\beta^0 = 0,\theta^0 = (d^Td)^{-1}d^Ty,z^0=0,t^0=0$.\\
{\textbf{for $k=0,\dots,K$ do}
\begin{enumerate}
\item $\theta^{k+1} = (d^Td)^{-1}d^T(y - X\beta^k)$. \label{alg2-step1}
\item $z^{k+1}  =z^{k} + \frac{\alpha}{m} X^T(y - d\theta^k - X\beta^k)$.\label{alg2-step2}
\item $\beta^{k+1} = \kappa\mathbf{Shrinkage}(z^{k+1})$
\item $t^{k+1} = (k+1)\alpha$.
\end{enumerate}
\textbf{end for}}\\
\textbf{Output:} Solution path $\{t^k, \theta^k,\beta^k\}_{k= 0,1,\dots,K}$.
\end{algorithm}


\subsection{Early Stopping Regularization} \label{sec:cv}
The Alg.\ref{alg-LBI-LME} actually returns a solution path with many estimators of different sparsity. So we need to find an optimal stopping time among $t^k=\alpha k$ to choose some best estimators and avoid overfitting. Here we sketch the procedure of cross-validation to choose the optimal stopping time:
\begin{itemize}
\item Given the training data, fix $\kappa$ and $\alpha$, then split the data into $K$ folds. Then choose a list of parameter $t$.
\item \textbf{for $k=1,\dots,K$ do}
	\begin{enumerate}
	\item Run Alg.\ref{alg-LBI-LME} on the training data except $k$-th fold to get the solution path.
	\item For pre-decided parameter list of $t$, use a linear interpolation to get $(\theta(t),\beta(t))$.
	\item On the $k$-th fold of training data, use the estimator $(\theta(t),\beta(t))$ to predict, and then compute prediction error.
	\end{enumerate}
	\textbf{end for}
\item  Return the optimal $t_{cv}$ with minimal average prediction error.
\end{itemize}
\textbf{Remark:} Because the Alg.\ref{alg-LBI-LME} only returns the estimator at discrete $\{t^k\}$ and may not contain the pre-decided parameter $t$, so we use a linear interpolation of the nearest two estimator $(\theta^{k},z^{k})$ and $(\theta^{k+1},z^{k+1})$ to approximate $(\theta(t),z(t))$. $\beta(t)$ is further obtained by using $\mathbf{Shrinkage}(z(t))$.

\begin{figure*}[t]
 \begin{center}
  \subfigure[Annotators with personalized ranking]{
\includegraphics[width=0.45\textwidth]{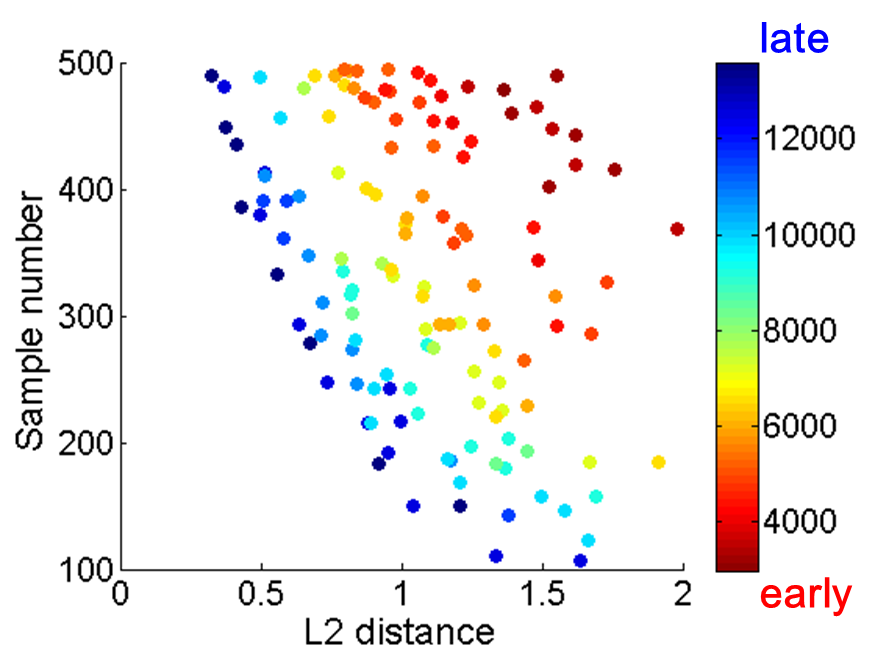}\label{simulated1}}
   \subfigure[Position-biased annotators]{
\includegraphics[width=0.45\textwidth]{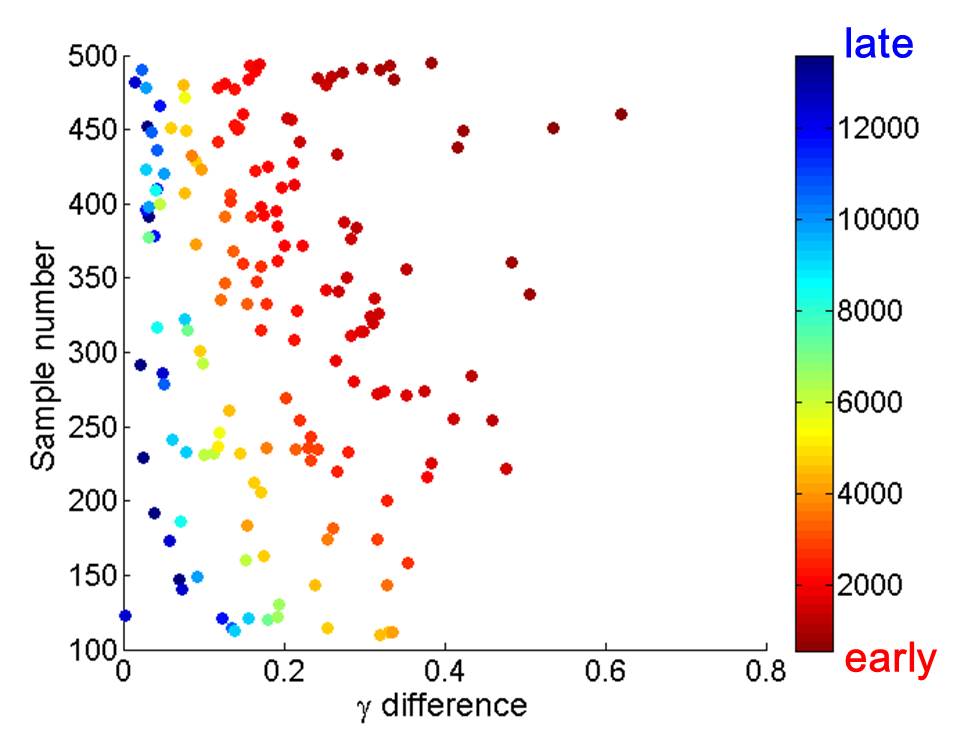}\label{simulated2}}
  \caption{Annotators with personalized ranking and position bias detected in simulated data.}
\end{center}
\end{figure*}

\section{Experiments}\label{sec:experiment}

In this section, four examples are exhibited with both
simulated and real-world data to illustrate the validity of
the analysis above and applications of the methodology proposed.
The first example is with simulated data while the
latter three exploit real-world data collected by crowdsourcing.

\subsection{Simulated Study}

\textbf{Settings} We validate the proposed algorithm on simulated data labeled by 500 annotators.
Specifically, we first generate the true $\theta_i \sim \N(0,1)$.
Then each annotator has a probability $p_1 = 0.4$ having a nonzero $\gamma^u$ and a probability $p_2 = 0.4$ having a nonzero $\delta^u$.
Those nonzero $\gamma^u$ is drawn randomly from $\N(0,0.2^2)$. And each entry $\delta_i^u$ of nonzero $\delta^u$ is drawn randomly from $\N(0,s^2)$ with $s \sim \mathbb{U}(0,0.3)$.
The noise $\varepsilon_{ij}^u$ is i.i.d. $\N(0,0.3^2)$. At last, we draw $N^u$ samples for each user randomly following the model \eqref{eq:linear}. The sample number $N^u$ uniformly spans in $[N_1,N_2] = [100,500]$. Here we choose $n=|V|=30$, which is consistent with the first real-world dataset.  Finally, we obtain a multi-edge graph with 150,494 pairwise comparisons annotated by 500 annotators.

\textbf{Results} Fig.\ref{simulated1} shows the annotators exhibiting personalized preferences selected via cross-validation, where each color dot represents one annotator. The optimal $t$ obtained via cross-validation is shown in Fig.\ref{simulated3}. The scores derived from each user are denoted as $\hat{\theta}^u$ and the common ranking as $\hat{\theta}$. Here the scores are the least squares solution of Eq.(\ref{eq:linear}). The X-axis represents the $L_2$-distance between each user's personalized ranking and the common ranking, $\|\hat{\theta}^u-\hat{\theta}\|$. The Y-axis counts the number of pairwise comparisons each user provides.
Clearly one can see the larger the $L_2$-distance and sample size, the more earlier this user is treated as one with personalized ranking (from color red to blue). This indicates that users jumped out earlier are those with large deviation from the population's opinion and a big sample size indicating a high confidence. Similar results of position-biased user is illustrated in Fig.\ref{simulated2}, in which the X-axis ($\gamma$ difference) represents the absolute difference of $\gamma$ between each user and the population.

Finally, to see whether our proposed method could provide more precise preference function for users by introducing individual-specific parameters (i.e., $\delta$ and $\gamma$), we split the data into training set ($70\%$ of the total pairwise comparisons) and testing set (the remaining $30\%$). To ensure the statistical
stability, we repeat this procedure 20 times. Tab.\ref{simulatedresult} shows the experimental results of the proposed mixed-effects model compared with HodgeRank, which indicates that our method exhibits smaller test error with an average of
$0.0948\pm 0.0008$ (in contrast to $0.1298\pm 0.0008$ in HodgeRank) due to its
parsimonious property.

\begin{figure}[t]
 \begin{center}
\includegraphics[width=0.7\linewidth]{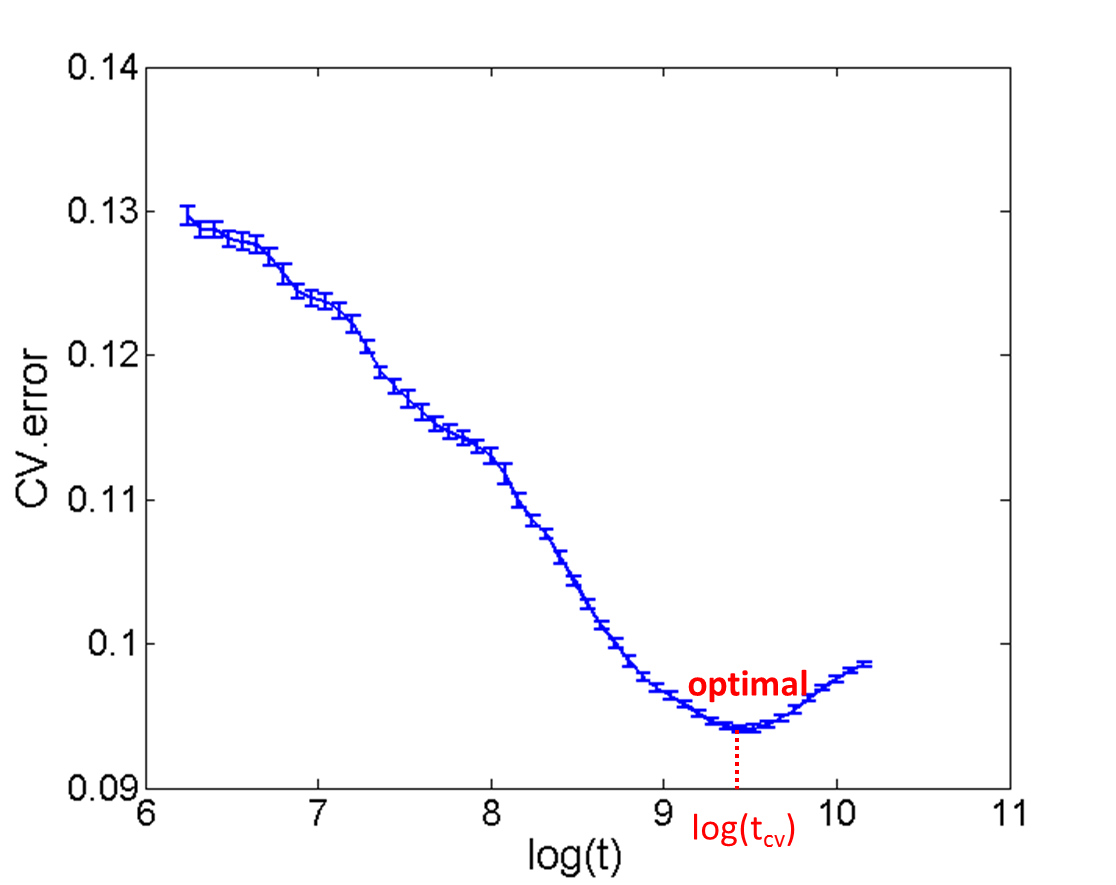}
  \caption{Optimal $t$ with minimal average prediction error in simulated data.} \label{simulated3}

\end{center}
\end{figure}

\begin{table}\caption{\label{simulatedresult} HodgeRank vs. Mixed-effects model in simulated data.}
\centering
\begin{tabular}{lllll}
 \hline     &min  &mean &max &std\\
 \hline  HodgeRank   & 0.1282   & 0.1298    &  0.1315      & 0.0008  \\
\hline  Mixed effects model   & 0.0934    & 0.0948    & 0.0961     & 0.0008  \\
 \hline
 \end {tabular}
\end{table}

\subsection{Human Age}

\begin{figure}[h]
 \begin{center}
\includegraphics[width=0.8\linewidth]{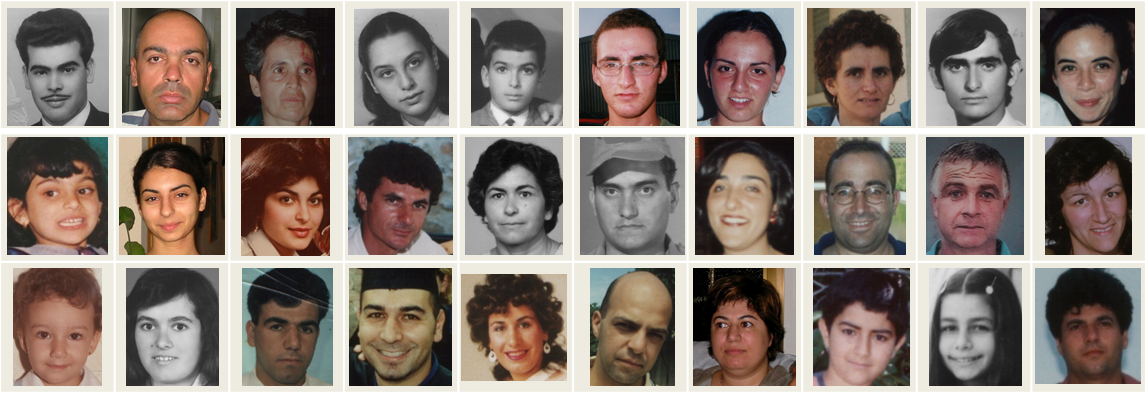}
  \caption{Images in Human age dataset.} \label{agedataset}
\end{center}
\end{figure}

\begin{table}[h]\caption{\label{tab:age} HodgeRank vs. Mixed-effects model in Human age dataset.}
\centering
\begin{tabular}{lllll}
 \hline     &min  &mean &max &std\\
 \hline  HodgeRank    &0.5542            &0.5716           &0.5907            &0.0101  \\
\hline  Mixed effects model  &0.4199             &0.4455           &0.4680           &0.0111   \\
 \hline
 \end {tabular}
\end{table}

\begin{figure}[h]
 \begin{center}
  \subfigure[LBI regularization path of $\delta$. (Red: top 10)]{
\includegraphics[width=0.26\textwidth]{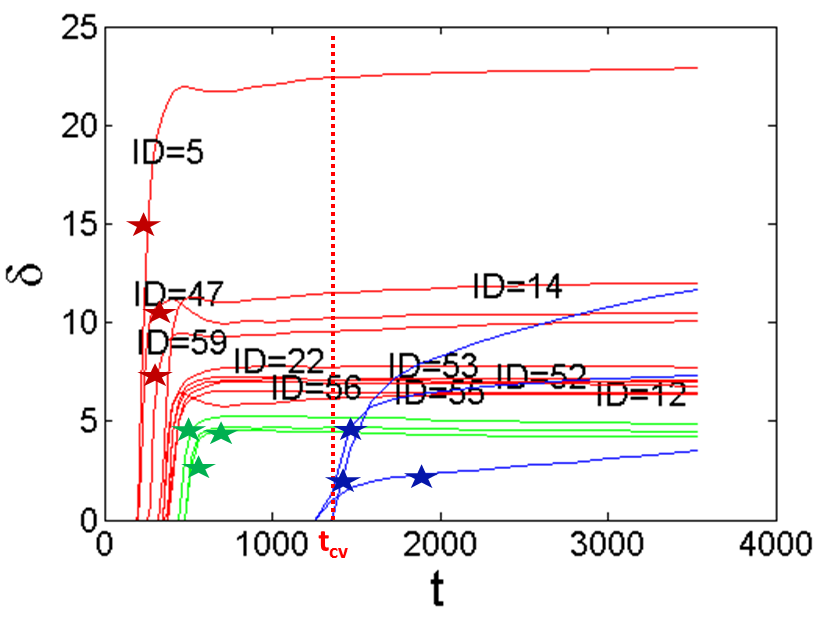}\label{fig:agepreferencepath}}
   \subfigure[Jumping out order.]{
\includegraphics[width=0.18\textwidth]{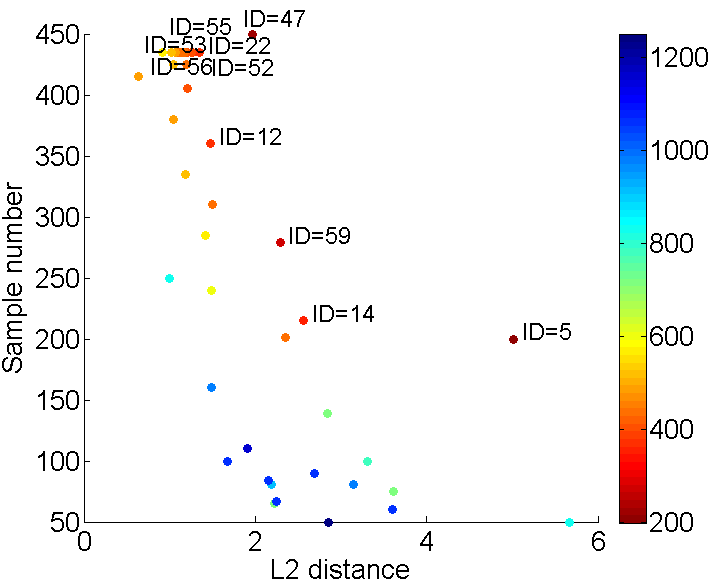}\label{fig:agepreferencedot}}
  \caption{Top 10 annotators with personalized ranking in Human age dataset.}
\end{center}
\end{figure}

\begin{figure}[h]
 \begin{center}
\includegraphics[width=0.75\linewidth]{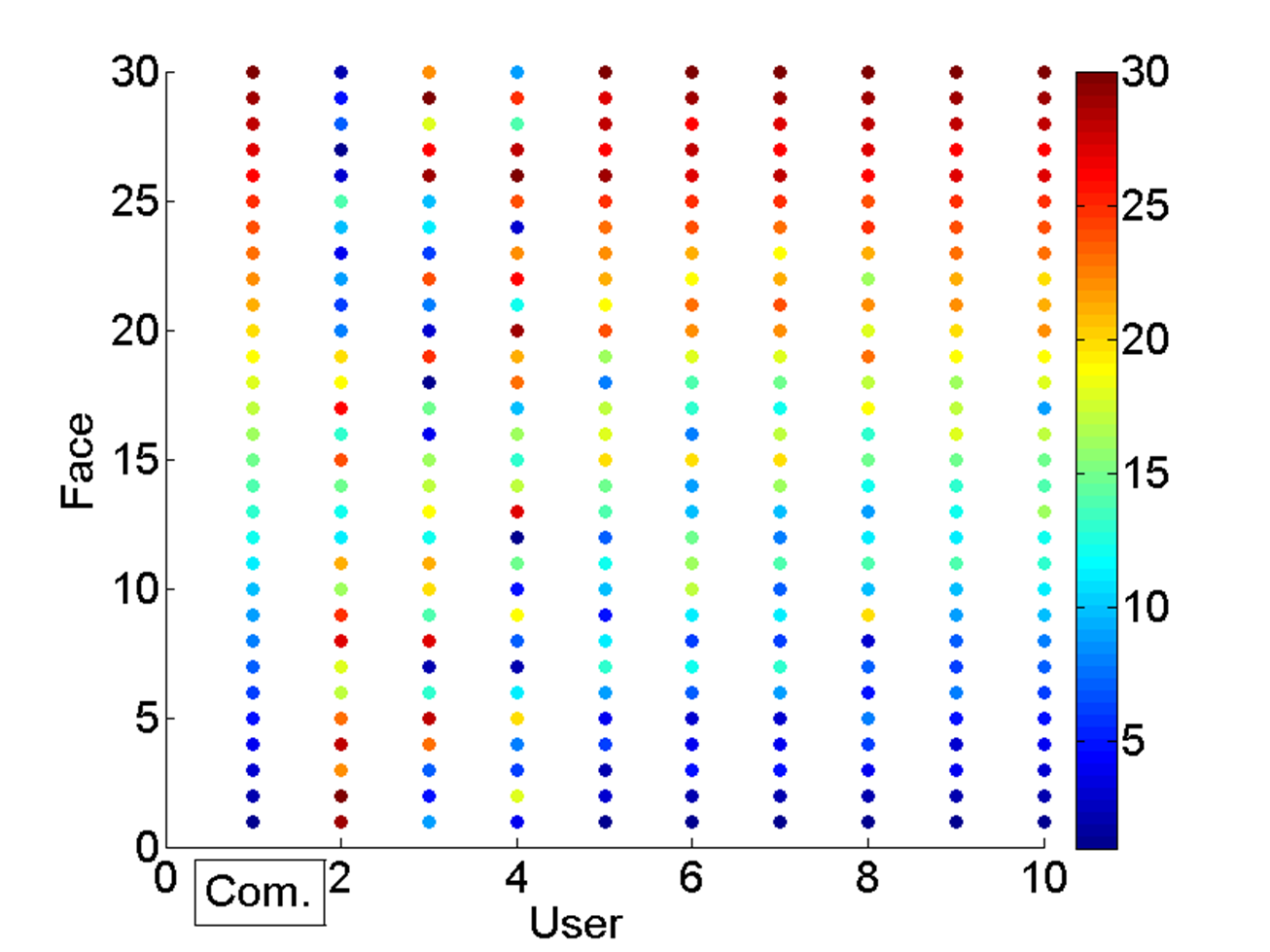}
  \caption{Comparison of common vs. personalized rankings of 9 representative annotators in Human age dataset.} \label{age_position_color}
\end{center}
\end{figure}

\begin{figure}[h]
 \begin{center}
\includegraphics[width=0.75\linewidth]{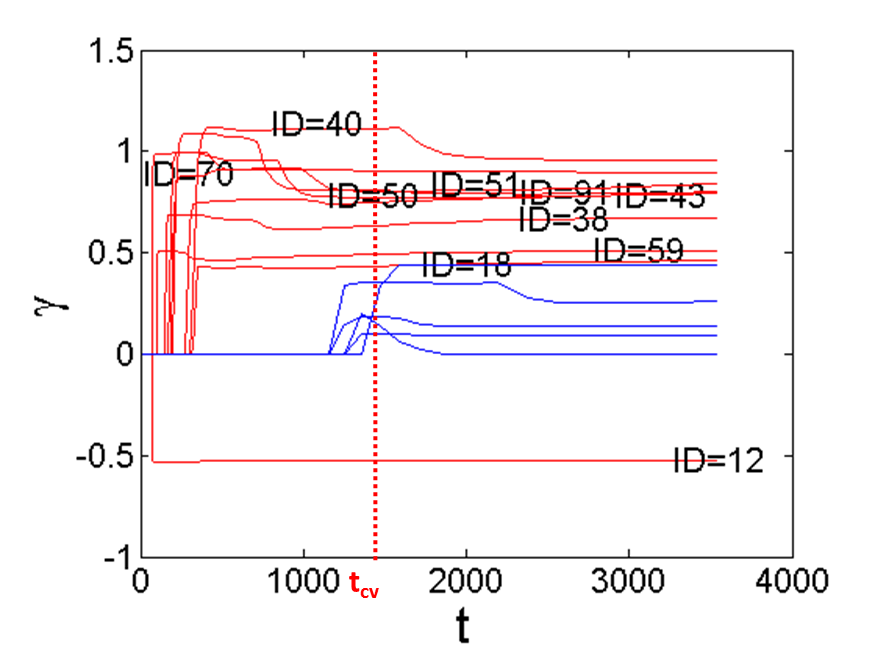}
  \caption{LBI regularization path of $\gamma$ in Human age dataset. (Red: top 10 position-biased annotators; Blue: bottom 5 position-biased annotators)} \label{fig:ageposition}
\end{center}
\end{figure}

{\renewcommand\baselinestretch{1.0}\selectfont
\setlength{\belowcaptionskip}{3pt}
\begin{table}[htb]\caption{\label{tab:ageposition} Top 10 position-biased annotators in Human age dataset, together with the click counts of each side (i.e., Left and Right).}
\tiny
\centering
\newsavebox{\tablebox}
\begin{lrbox}{\tablebox}
\begin{tabular}{||c|c|c|c||c|c|c|c||}
  \hline  \textbf{Order} &\textbf{ID}   &\textbf{Left}  &\textbf{Right} & \textbf{Order} & \textbf{ID}   &\textbf{Left}  &\textbf{Right} \\
 \hline
\hline    1 & \textbf{12}	&90	&270  & 6 & \textbf{91}	&79	&5\\
\hline   2 & \textbf{70}	&191	&9 & 7 & \textcolor{blue}{\textbf{51}}	&63	&0\\
\hline    3 & \textbf{59}	&213	&66 & 8 & \textbf{50}	&60	&3 \\
\hline    4 & \textbf{38}	&110 &15  & 9 &\textbf{18}	&74	&25 \\
\hline    5 & \textbf{43}   &79 & 1 & 10 & \textcolor{blue}{\textbf{40}}	&40	&0   \\

\hline
\end{tabular}
\medskip
\end{lrbox}
\scalebox{1}{\usebox{\tablebox}}
\end{table}
\par}

\textbf{Settings} In this dataset, 30 images from human age dataset FG-NET \footnote{http://www.fgnet.rsunit.com/} are annotated by a group of volunteer users on \href{http://www.chinacrowds.com/}{ChinaCrowds} platform, as is illustrated in Fig.\ref{agedataset}. The annotator is presented with two images and
given a binary choice of which one is older. Totally, we obtain 14,011 pairwise comparisons from 94 annotators.

\textbf{Results} Tab.\ref{tab:age} shows the mean test error (70\% data for training, 30\% for testing)
results of 20 times achieved by this scheme. It is shown that this mixed-effects model could provide
better approximate results of the annotators' preference than the HodgeRank estimator, with an
average test error of $0.4455\pm 0.0111$ (in contrast to $0.5716\pm 0.0101$ in HodgeRank).\\

To further investigate the characteristics of annotators with personalized ranking, Fig.\ref{fig:agepreferencepath} illustrates annotator's LBI regularization paths of preference deviations with optimal $t$ (i.e., $t_{cv}$) returned by cross-validation, while Fig.\ref{fig:agepreferencedot} shows the relationships among $L_2$-distance to the common ranking, sample number and jumping out order of each annotator.  The red curves in Fig.\ref{fig:agepreferencepath} represent the top 10 annotators who jumped out early which are also marked in Fig.\ref{fig:agepreferencedot}. Similar to the simulated data, annotators jumped out earlier are those with a large deviation ($L_2$-distance) from the common ranking and a big sample size showing high confidence of such deviations. Moreover, Fig.\ref{age_position_color} shows the order comparisons of common ranking (i.e., com.) and personalized ranking of 9 representative annotators at $t_{cv}$. The X-axis represents user index: user = 2, 3, 4 jumped out early corresponding to paths labeled with red stars in Fig.\ref{fig:agepreferencepath}; user = 5, 6, 7 jumped out in the middle time corresponding to green stars; user = 8, 9, 10 jumped out late corresponding to blue stars. The order of faces in Y-axis is arranged from lower to higher (i.e., from color blue to red) according to the common ranking score calculated by our method. The color represents the ranking position returned by the corresponding user. It is easy to see users jumped out late exhibit almost consistent ranking order with the common ranking, while the earlier ones are almost the adversarial against the common. A subset of this has been shown in Fig.\ref{fig:age0} in introduction.

Moreover, Fig.\ref{fig:ageposition} illustrates the LBI regularization paths of annotator's position bias with red lines represent the top 10 annotators. Tab.\ref{tab:ageposition} further shows the click counts of each side (i.e., Left and Right) for these top 10 position-biased annotators.
It is easy to see that these annotators can be divided
into two types: (1) click one side all the time (with
ID in blue); (2) click one side with high probability
(others). Although it
might be relatively easy to identify the annotators of type (1) above
by inspecting their inputs, it is impossible for eye inspection
to pick up those annotators of type (2) with mixed rational and abnormal behaviors.
Therefore it is essential to design such a statistical methodology to
quantitatively detect these kind of position-biased annotators
for crowdsourcing platforms in market.

\begin{figure}[t]
 \begin{center}
\includegraphics[width=0.8\linewidth]{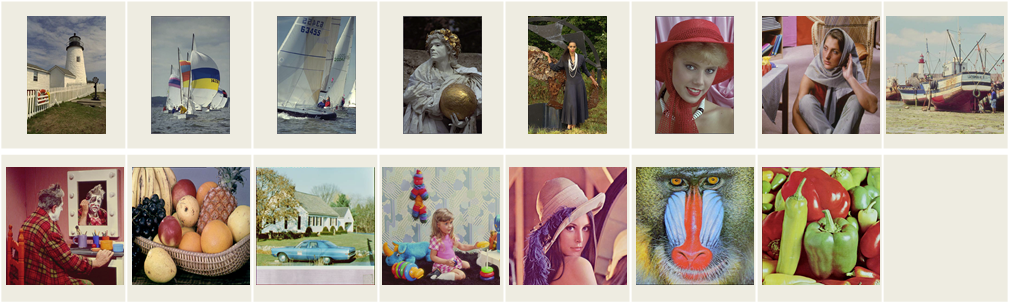}
  \caption{Images in LIVE and IVC dataset.} \label{iqadataset}
\end{center}
\end{figure}

\subsection{Image Quality Assessment (IQA)}

\textbf{Settings} Two publicly available datasets, LIVE
\cite{LIVE} and IVC \cite{IVC}, are used in this work. The LIVE dataset contains 29 reference images and 779 distorted images. Considering the resolution limit of most test computers, we only choose 6 different reference images ($480 \times 720$) and 15 distorted versions of each reference, for a total of 96 images. The second dataset, IVC, which is also a broadly adopted dataset in the community of multimedia quality evaluation, includes 10 reference images and 185 distorted images derived from four distortion types--JPEG2000, JPEG, LAR Coding, and Blurring. Following the collection strategy in LIVE, we further select 9 different reference images ($512 \times 512$) and 15 distorted images of each reference. Totally, we obtain a medium-sized image set that contains a total of 240 images from 15 references, as illustrated in Fig.\ref{iqadataset}.
Finally, 342 observers of different
cultural background, each of whom performs a varied number of comparisons via Internet, provide
$52,043$  paired comparisons in total. The number of responses each reference image receives is different.

\begin{table}[t]\caption{\label{tab:iqa} HodgeRank vs. Mixed-effects model in IQA dataset (reference image 1).}
\centering
\begin{tabular}{lllll}
 \hline     &min  &mean &max &std\\
 \hline  HodgeRank    &0.4918            & 0.5135           &0.5429            &0.0134  \\
\hline  Mixed effects model  &0.2922           & 0.3241         &0.3576          & 0.0158   \\
 \hline
 \end {tabular}
\end{table}

\begin{figure}[t]
 \begin{center}
  \subfigure[LBI regularization path of $\delta$. (Red: top 10)]{
\includegraphics[width=0.26\textwidth]{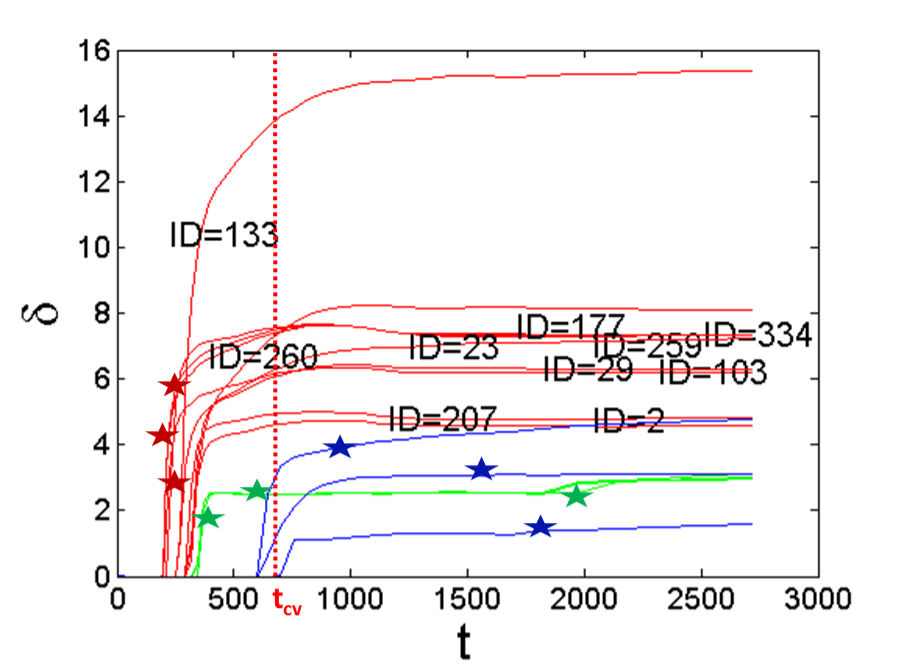}\label{fig:iqaref1preferencepath}}
   \subfigure[Jumping out order.]{
\includegraphics[width=0.18\textwidth]{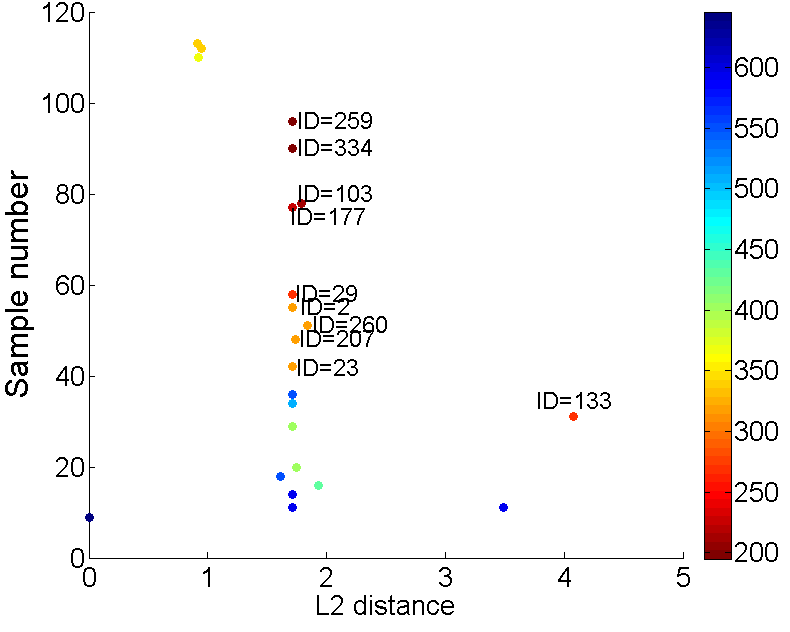}\label{fig:iqaref1preferencedot}}
  \caption{Top 10 annotators exhibiting personalized ranking in IQA dataset (reference image 1).}
\end{center}
\end{figure}

\begin{figure}[htb]
 \begin{center}
\includegraphics[width=0.65\linewidth]{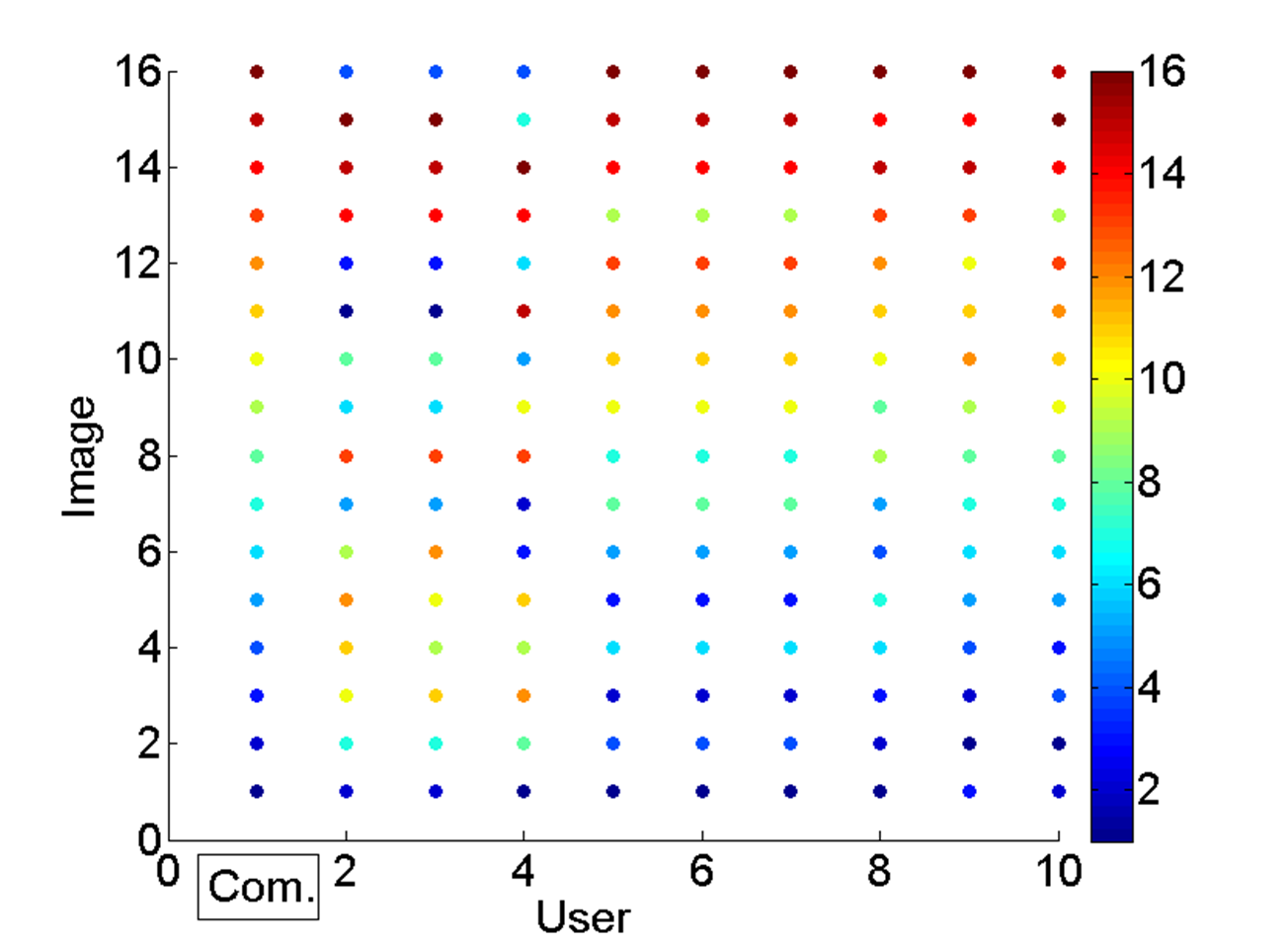}
  \caption{Comparison of common vs. personalized rankings of 9 representative annotators in IQA dataset (reference image 1).} \label{iqa_position_color}
\end{center}
\end{figure}

\begin{figure}[htb]
 \begin{center}
\includegraphics[width=0.7\linewidth]{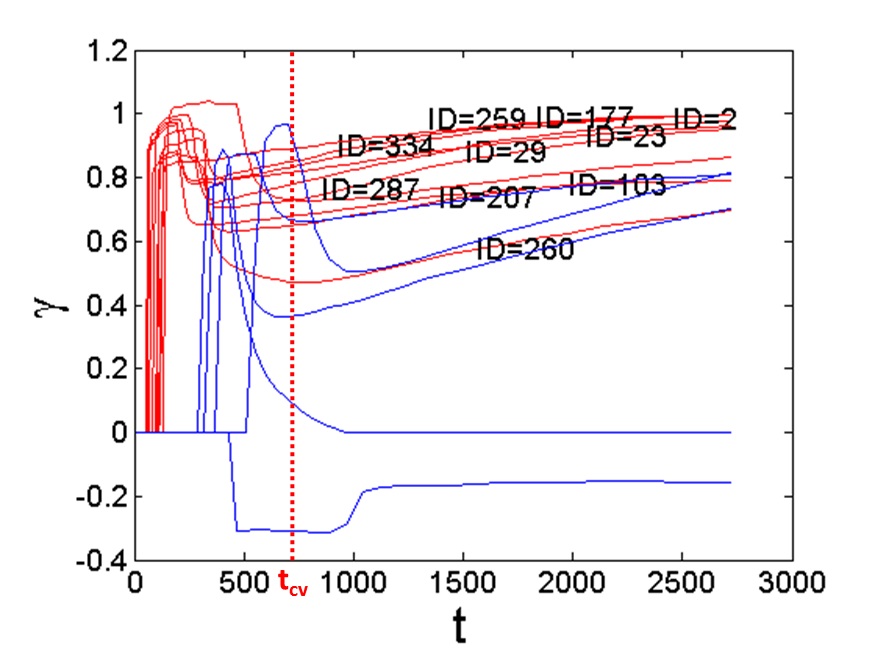}
  \caption{LBI regularization path of $\gamma$ in IQA dataset (reference image 1). (Red: top 10 position-biased annotators; Blue: bottom 5 position-biased annotators).} \label{fig:iqa_position}
\end{center}
\end{figure}

{\renewcommand\baselinestretch{1.0}\selectfont
\setlength{\belowcaptionskip}{0pt}
\begin{table}[htb]\caption{\label{tab:ref1} Top 10 position-biased annotators in IQA dataset (reference image 1).}
\tiny
\centering
\begin{lrbox}{\tablebox}
\begin{tabular}{||c|c|c|c||c|c|c|c||}
  \hline  \textbf{Order} &\textbf{ID}   &\textbf{Left}  &\textbf{Right} & \textbf{Order} & \textbf{ID}   &\textbf{Left}  &\textbf{Right} \\
 \hline
 \hline  1 & \textcolor{blue}{\textbf{259}}    &96     &0  & 6 & \textcolor{blue}{\textbf{2}}   &55     &0 \\
\hline  2 & \textcolor{blue}{\textbf{334}}     &90    & 0  & 7 & \textbf{260}    & 49     &2\\
 \hline  3 & \textcolor{blue}{\textbf{177}}   &77     &0   & 8 & \textcolor{blue}{\textbf{23}}    &42     &0 \\
 \hline 4 & \textbf{103}  &74     &4  &9  & \textbf{207}    &46     &2\\
 \hline 5 & \textcolor{blue}{\textbf{29}}  &58     &0  & 10 & \textcolor{blue}{\textbf{287}}   &34     &0\\

 \hline
\end {tabular}
\medskip
\end{lrbox}
\scalebox{1}{\usebox{\tablebox}}
\end{table}
\par}
%

To validate whether the annotators' preference function we estimated is good enough, we randomly take
reference image 1 as an illustrative example while other reference
images exhibit similar results.

\textbf{Results} In terms of prediction performance, it can be seen that the mixed-effects model is more accurate than HodgeRank by significantly reducing the mean test error, as is shown in Tab.\ref{tab:iqa}.

Besides, Fig.\ref{fig:iqaref1preferencepath} and \ref{fig:iqaref1preferencedot} shows the regularization paths of personalized ranking with top 10 annotators (red curves) selected early in the paths, as well as comparisons in order of jumping, sample size, and common ranking.
It is easy to see that among these 10 annotators, 9 of them (except annotator with ID = 133) exhibit almost the same $L_2$-distance with the common ranking. The reason behind this
is these 9 annotators click one side all the time (i.e., position-biased annotators), thus inducing a large $\gamma$. In this case, each image quality scores estimated from each annotator are close to 0 (though have differences), thus deriving the almost the same $L_2$-distance with the common ranking scores. Similar to the Age dataset, the common ranking vs. personalized ranking results of 9 representative users is shown in Fig.\ref{iqa_position_color}, corresponding to the red/green/blue stars in Fig.\ref{fig:iqaref1preferencepath}.

Moreover, the LBI regularization paths of position bias $\gamma$ and click counts of top 10 annotators in this dataset are shown in Fig.\ref{fig:iqa_position} and Tab.\ref{tab:ref1}. It is easy to see that annotators picked out are mainly those clicking on
one side almost all the time. Besides, it is interesting to see
that all these annotators highlighted with blue color in
Tab.\ref{tab:ref1} click the left side all the time. We then go back to
the crowdsourcing platform and find out that the reason behind
this is a \emph{default choice} on the left button, which induces
some lazy annotators to cheat for the task.

\subsection{WorldCollege Ranking}

\textbf{Settings} We now apply the proposed method to the WorldCollege dataset, which is composed of 261 colleges. Using the \href{http://www.allourideas.org/}{Allourideas} crowdsourcing platform, a total of 340 distinct annotators from various countries (e.g., USA, Canada, Spain, France, Japan)
are shown randomly with pairs of these colleges, and asked to decide which of the
two universities is more attractive to attend. Finally, we obtain a
total of 8,823 pairwise comparisons.

\textbf{Results} We apply the proposed method to
the resulting dataset and find out that, similar to the simulation and other two real-world datasets, the mixed-effects model could produce better performance than Hodgerank with smaller mean test error, shown in Tab.\ref{tab:university}. Noting that in this dataset, only 7 annotators are treated as annotators with distinct personalized rankings at optimal $t$ (i.e., $t_{cv}$) selected via cross-validation, as is shown in Fig.\ref{fig:universitypreferencepath} and \ref{fig:universitypreferencedot}.
The reason why others with relative big $\delta$ are not detected out lies in the small sample size they provide indicating high variances. The common ranking vs. personalized ranking of these 7 users is shown in Fig.\ref{university_position_color} with a similar observation to the other datasets.
Besides, the regularization paths of position bias and click counts of top 10 annotators in this dataset are shown in Fig.\ref{fig:university_position} and Tab.\ref{tab:universityposition}.
It is easy to see that similar to the human age dataset, these annotators are either clicking one
side all the time, or clicking one side with high probability in mixed behaviors.
%

\begin{table}[htb]\caption{\label{tab:university} HodgeRank vs. Mixed-effects model in WorldCollege ranking dataset.}
\centering
\begin{tabular}{lllll}
 \hline     &min  &mean &max &std\\
 \hline  HodgeRank     & 0.8089    & 0.8217     & 0.8406     & 0.0078 \\
\hline  Mixed effects model   & 0.7066     & 0.7199     & 0.7398    &  0.0088   \\
 \hline
 \end {tabular}
\end{table}

\begin{figure}[htb]
 \begin{center}
  \subfigure[LBI regularization path of $\delta$. (Red: top 7; Blue: others)]{
\includegraphics[width=0.26\textwidth]{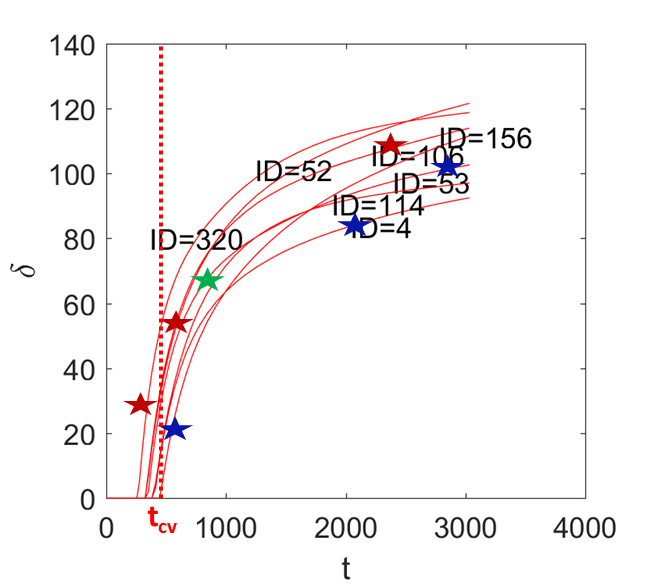}\label{fig:universitypreferencepath}}
   \subfigure[Jumping out order.]{
\includegraphics[width=0.18\textwidth]{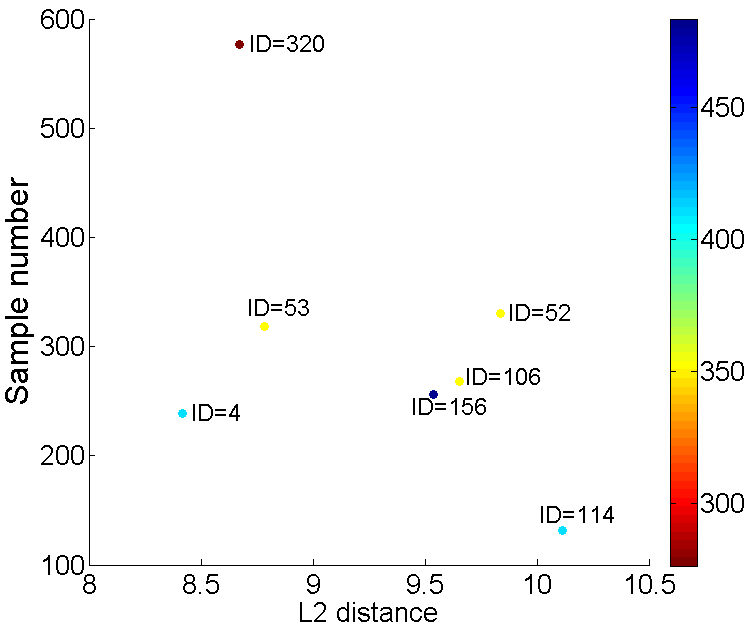}\label{fig:universitypreferencedot}}
  \caption{The 7 annotators with personalized ranking in WorldCollege ranking dataset.}
\end{center}
\end{figure}

\begin{figure}[htb]
 \begin{center}
\includegraphics[width=0.62\linewidth]{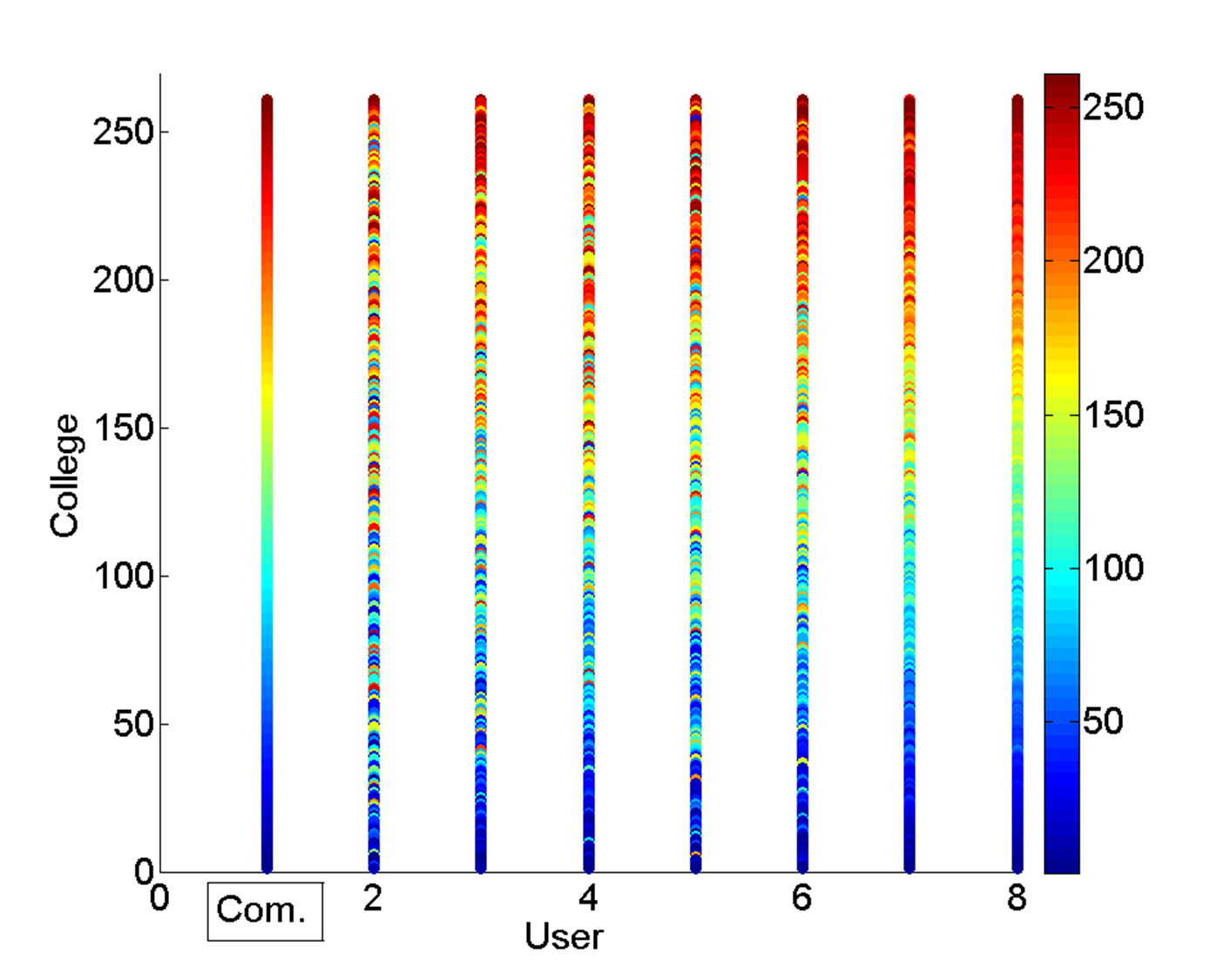}
  \caption{Comparison of common vs. personalized rankings of 7 annotators in WorldCollege ranking dataset.} \label{university_position_color}
\end{center}
\end{figure}

{\renewcommand\baselinestretch{1.0}\selectfont
\setlength{\belowcaptionskip}{0pt}
\begin{table}[htb]\caption{\label{tab:universityposition} Top 10 position-biased annotators in WorldCollege ranking dataset.}
\tiny
\centering
\begin{lrbox}{\tablebox}
\begin{tabular}{||c|c|c|c||c|c|c|c||}
  \hline  \textbf{Order} &\textbf{ID}   &\textbf{Left}  &\textbf{Right} & \textbf{Order} & \textbf{ID}   &\textbf{Left}  &\textbf{Right} \\
 \hline
 \hline  1 & \textcolor{blue}{\textbf{268}} &148	&0 & 6 & \textbf{270}	&20	&70 \\
 \hline  2 & \textcolor{blue}{\textbf{209}}	&127	&0 & 7 & \textcolor{blue}{\textbf{267}}	&45	&0 \\
 \hline 3 & \textbf{156}	&189	&67 & 8 & \textbf{276}	&16	&54 \\
 \hline 4  & \textbf{320}	&253	&324  & 9 & \textcolor{blue}{\textbf{166}}	&35	&0 \\
 \hline 5 & \textbf{87}	&11	&62 & 10 & \textcolor{blue}{\textbf{115}}	&34	&0 \\
 \hline
\end {tabular}
\medskip
\end{lrbox}
\scalebox{1}{\usebox{\tablebox}}
\end{table}
\par}

\begin{figure}[h]
 \begin{center}
\includegraphics[width=0.72\linewidth]{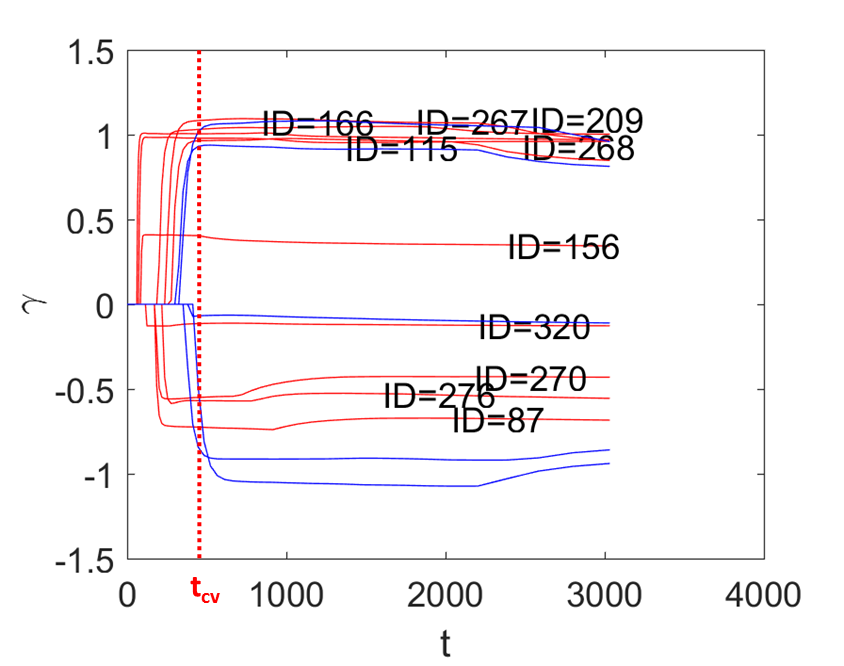}
  \caption{LBI regularization path of $\gamma$ in WorldCollege ranking dataset. (Red: top 10 position-biased annotators; Blue: bottom 5 position-biased annotators).} \label{fig:university_position}
\end{center}
\end{figure}


\section{Conclusions}\label{sec:conclusion}

In this paper, we propose a parsimonious mixed-effects model based on HodgeRank to learn user's preference or utility function in crowdsourced ranking, which
takes into account both the personalized preference deviations from the common and position biases of the annotators.
To be specific, common preference scores indicate the consistent ranking on population-level which approximates the behavior of
all users, while a small set of annotators might have nonzero
personalized deviations and abnormal behavior in position bias. Equipped with the newly developed Linearized
Bregman Iteration, which is a simple iterative procedure generating a sequence of parsimonious models,  we establish a dynamic path from the common utility to individual
variations, with different levels of parsimony or sparsity on personalization.
Experimental studies are conducted with both simulated examples and real-world
datasets. Our results suggest that the proposed methodology is an effective tool to investigate the diversity in annotator's behavior
in modern crowdsourcing data.

\section*{Acknowledgements}

The research of Qianqian Xu was supported by National Key Research and Development Plan (No. 2016YFB0800603), National Natural Science Foundation of China (No. 61422213, 61402019, 61390514, 61572042), ``Strategic Priority Research Program" of the Chinese Academy of Sciences (XDA06010701), National Program for Support of Top-notch Young Professionals, and China Postdoctoral Science Foundation (2015T-80025).
The research of Jiechao Xiong and Yuan Yao
was supported in part by National Basic Research Program of China under
grant 2015CB85600, 2012CB825501, and NSFC grant 61370004, 11421110001
(A3 project), as well as grants from Baidu and Microsoft Research-Asia. We would like to
thank anonymous reviewers who gave
valuable suggestions to help improve the manuscript.

\bibliographystyle{abbrv}

\bibliography{sigproc}  

\end{document}